\documentstyle[epsfig]{elsart}
\begin{document}
\begin{frontmatter}
\title{On the different role of protons and neutrons in 
antinucleon annihilations on nuclei.} 
\author[Brescia]{A.~Bianconi},
\author[Brescia]{G.~Bonomi},
\author[Brescia]{M.P.~Bussa},
\author[Brescia]{G.~Gomez}, 
\author[Brescia]{E.~Lodi Rizzini},
\author[Brescia]{L.~Venturelli},
\author[Brescia]{A.~Zenoni},
\address[Brescia]{Dip. di Chimica e Fisica per 
l'Ingegneria e per i Materiali, 
Universit\`{a} di Brescia and
INFN, Sez. di Pavia, Italy} 

\begin{abstract} 
We compare data of antineutron and antiproton annihilation 
cross sections on different targets at very low energies. 
After subtracting Coulomb effects, we observe that 
the ratio between the $\bar{n}p$ and $\bar{p}p$ 
annihilation cross sections is an oscillating function 
of the energy at momenta smaller 300 MeV/c. 
This nontrivial behavior 
is confirmed by the analysis of the relative number of 
$\bar{p}n$ and $\bar{p}p$ annihilations in nuclei. 
We show that a part of the strong shadowing 
phenomena in $\bar{p}$-nucleus annihilations can be explained 
in terms of this oscillation, while a part requires 
different explainations. 
\end{abstract} 
\end{frontmatter} 

\section{Introduction.} 

Recently data on $\bar{n}p$ annihilation in the range 
40-400 MeV/c (for the laboratory $\bar{n}$ momentum $k$) 
have been produced by the Obelix Collaboration\cite{feli}. 
We would like to compare these with other data on 
$\bar{p}p$ and $\bar{N}$-nucleus annihilation. In particular, 
we are interested in: $\bar{n}$ 
annihilation on nuclei from $^{12}$C to $^{207}$Pb in the 
range 180-280 MeV/c\cite{ableev}; $\bar{p}p$ annihilation from 
30 to 180 MeV/c\cite{obe1}; $\bar{p}$D, $\bar{p}^4$He and 
$\bar{p}^{20}$Ne at small momenta, down 
to 45 MeV/c\cite{obe2,ne991}; 
$\bar{p}$-nucleus annihilation on intermediate nuclei at 
larger momenta (over 200 MeV/c, see\cite{bende} for 
a recollection of these data); 
and on the ratio between 
$\bar{p}p$ and $\bar{p}n$ annihilations inside nuclear 
targets\cite{nb2}. We will start by comparing $\bar{p}p$ 
and $\bar{n}p$ data, and then we will try to correlate 
these with the low energy nuclear data. 

One of our aims is simply to compare $\bar{n}p$ 
and $\bar{p}p$ annihilation cross sections. This comparison 
is a delicate operation, because it requires 
subtraction of Coulomb effects, and comparison of data coming 
from different experiments with antiproton and antineutron 
beams. Due to the difficulties in calculating a 100 \% 
reliable flux normalization, it is preferable, as far as possible, 
to compare data from the same experiment. 

A second goal of this work is to 
establish which is the degree of correlation between 
antineutron-nucleon and antinucleon-nucleus data. 
Indeed, as better discussed later, $\bar{p}p$ data below 
600 MeV/c can be well fitted via energy independent 
optical potentials, while $\bar{n}p$ data show a 
nontrivial energy dependence that should be reflected in 
the energy dependence of nuclear annihilations. 

Another interesting point is the role of the different 
isospin channels in the low energy nuclear shadowing. 
It has been demonstrated that a marked nuclear shadowing  
characterizes low energy $\bar{N}$-nucleus annihilation cross 
sections\cite{obe2,obe4,wid,wyc1,wyc2,pro1,pro2,pro3,n1,n2,n3,n4}. 
For this phenomenon it is possible to imagine two classes of 
explainations. First, general 
quantum mechanical processes, as described in a later section. 
Mechanisms of this kind do not discriminate too much between 
different annihilation channels. 
On the other side we may imagine a role for peculiar properties 
of single nuclear species. Indeed, the data are rather 
incomplete and impossible to organize in a systematic way. 
In particular 
a different $N/Z$ composition, or a different $N/Z$ 
distribution near the nuclear surface can be relevant, since 
most of the theoretical models (for reviews see 
e.g. \cite{dov1,amsl,kerb,walch,wei93,eri88,gn84}) 
establish peculiar properties for the different isospin 
channels, 
properties which are normally derived either from G-parity 
(or C-parity\cite{dst94}) transformations of the better 
known nucleon-nucleon potentials, or from peculiar quark 
diagrams. 
So, to summarize this point, it is relevant to establish how 
much of the low energy nuclear shadowing of annihilation 
processes can be due to general mechanisms, and how much 
is due to the nuclear composition in terms of protons and 
neutrons, and to the details of the nuclear structure. 

We limit ourselves to a phenomenological 
analysis of the data, with standard methods of nuclear 
physics, without entering the debate about the 
underlying hadronic structure mechanisms governing the 
annihilation process. Of course, it is not always possible 
to separate completely the two levels of the analysis, i.e. 
nuclear and hadronic. Wide reviews on the debate 
concerning the annihilation mechanisms can be found in the 
references listed above. Experimental (scattering and 
annihilation) 
data up to 1994 are recollected in \cite{bende}, 
data on antiprotonic atoms up to 
1988 in \cite{bat90}, and  
a very recent account of the many problems related with the 
antinucleon experimental techniques can be found 
in \cite{eadh}. 

We still lack a well founded method that can be properly 
used at low energies to link the shadowing properties (that 
automatically arise from an optical potential 
treatment\cite{pro1,n1,pro3}) 
with the details of the nuclear structure and internal motions 
(which are taken into account in Impulse Approximation inspired 
treatments). The main 
exception in this respect is Deuteron, which has been 
considered by some authors\cite{wyc1,wyc2,pro3} and 
where it was possible to relate clearly the overall  
antinucleon-nucleus processes with the antinucleon-nucleon 
interactions. 
With more complex nuclei, optical potential analyses  
have been carried on to study antiprotonic 
atoms\cite{bat87,del74,thal85,rook79,gn87,dumb87}. At 
momenta over 200 MeV/c 
the KMT method\cite{kmt} has been used in \cite{aar} 
and the Glauber method\cite{glau} in \cite{aldo}
(and probably in other works). 
In the present work our hope is that some 
interesting information can be separately extracted 
from a PWIA and an optical potential analysis. 

For the sake of brevity we will indicate total 
$\bar{n}$ and $\bar{p}$ annihilation cross 
sections with TNA and TPA respectively, specifying the target. 
We will indicate as $F_A$ the ratio between $\bar{p}n$ 
and $\bar{p}p$ annihilations in a given nucleus with 
atomic number $A$. 

Concerning the choice of an optical potential for 
discussing nucleon-antinucleon annihilation, 
it is well known that many quite 
different optical potentials fitted 
successfully pre-Obelix nucleon-antinucleon 
data\cite{dst94,bp,dr1,rs82,paris,free,alb,paris2,mizu,mt85,tmm,kw86,bonn1} 
and they can probably fit the data under discussion too, 
since these data do not represent such a strong contraint. 
Here and in previous papers\cite{n1,n2,n3} we 
have relied on very simple optical potentials, with 
Woods-Saxon shape, the same parameters for each 
spin channel and no energy dependence in the 
momentum range 0-600 MeV/c (more indications are given in 
section 3).  Although such potentials 
cannot reproduce the full phenomenology of nucleon-antinucleon 
interactions, they seem sufficient 
for fitting the available $\bar{p}p$ low energy 
data. With the $\bar{n}p$ annihilation data, as 
described in section 3, a higher level of sophistication is 
perhaps necessary. 

\section{The role of Coulomb interactions. Apparent violation 
of the isospin invariance.}

Comparison of annihilation data with different projectiles 
and targets is nonsense without subtracting Coulomb effects 
at momenta below 200 MeV/c. The traditional  
estimation\cite{ll1} has undercome 
modifications\cite{cp1,cp2,n2} in the last years. 
Qualitatively the effect 
of the Coulomb forces is to focus the projectile wavefunction 
in the annihilation region, which is a spherical shell 
of thickness 0.5-1 fm\cite{dov1,brue91,aar} at 
the target proton/nucleus surface. 

In a previous work\cite{n2} we gave analytical expressions 
for the correcting $\bar{p}$-nucleus enhancement factors. 
E.g., for the $\bar{p}p$ annihilation cross section the 
enhancement factor can be reproduced within 5 \% in the 
range 30-400 MeV/c by the function $1+0.0003\beta^{-2}$. 
To calculate the enhancement factor we considered 
some completely different optical potentials. 
All of them included  
the electrostatic potential of a spherical charge 
distribution and fitted the available low energy data. 
Then we removed the electrostatic potential and calculated 
again the annihilation rates. The obtained Coulomb 
enhancement factor was sufficiently independent of 
the choice of the strong part of the optical potential, 
in the considered momentum range. 

An observation is necessary concerning 
apparent ``violations'' 
of the isospin invariance. Isospin symmetry suggests that: 
(i) $\bar{n}p$ cross sections are equal to $\bar{p}n$ ones; 
(ii) $\bar{p}p$ and $\bar{n}n$ are equal apart from 
electromagnetic effects; (iii) the four cross sections that 
one can imagine are actually combinations of two. In principle 
this is undoubtful, but in practice it does not work, and 
an example can clarify the point. Normally, $\bar{n}p$ 
data come from collisions between free antineutrons and 
protons, while $\bar{p}n$ 
cross sections are extracted from deuteron targets, with two 
consequences: 

1) The antiproton is attracted by the deuteron charge. The 
range of action of the Coulomb forces is much larger than the 
deuteron radius, with the result that the 
$\bar{p}n$ annihilation rate 
is almost as much ``Coulomb distorted'' as the 
$\bar{p}p$ one, so it should be expected to be larger (much 
larger at very low energies) than the $\bar{n}p$ one. This 
problem is not present at momenta $>>$ 100 MeV/c, where 
Coulomb focusing effects can be neglected. 

2) As already noticed\cite{obe2,wid,pro1,n1,pro2,n3},  
below 60 MeV/c 
shadowing effects in nuclei are so strong that the $\bar{p}$D 
annihilation cross section is smaller than the corresponding 
$\bar{p}p$ one in the laboratory frame, or approximately 
equal in the center of mass frame (due to the different 
transformations 
relating laboratory with center of mass variables for the cases 
of proton and deuteron targets).  
In both cases $\sigma_{\bar{p}D}$ is much smaller than 
$\sigma_{\bar{p}p}+\sigma_{\bar{n}p}$. 

The previous example should clarify that, 
although it is very likely that one would find  
$\sigma_{\bar{n}p}$ $\approx$ $\sigma_{\bar{p}n}$ 
in an experiment on $free$ neutrons, 
comparisons involving different isospin channels 
for $k$ $<<$ 100 MeV/c should be 
performed with the greatest care as far as free neutron targets,
or antiproton targets, are not available. One 
should therefore be aware that an isospin decomposition cannot be 
completely free from model dependence. 

\section{Nuclear shadowing effects.}

Recent experimental data\cite{obe2,wid,n3} show 
that at antinucleon momenta (in the laboratory) 
below 60 MeV/c the nuclear 
shadowing effect is very strong. Below 60 MeV/c 
antiproton annihilation rates on Deuteron and $^4$He 
are smaller than on Hydrogen. The annihilation 
rate on $^{20}$Ne is larger but not that 
much\cite{ne991}. This 
and related phenomena has been discussed by us and 
other authors\cite{wyc1,wyc2,pro1,pro2,n1,n4}. 

Rather independently of the mechanism underlying the 
annihilation process, it had been 
previously demonstrated that in the framework of the 
multiple scattering theory\cite{wyc1}, of variational 
methods\cite{wyc2} and of optical potential 
treatments\cite{pro1,n1} one can predict such shadowing 
effects. It has been reported\cite{pro1} that also in 
the coupled-channel approach one can obtain the same 
result.  

The fact that different methods lead to similar results
suggested us to investigate the problem from a more 
general and qualitative, although less precise, point 
of view. 
In our work\cite{n4} we have shown that 
due to the quantum uncertainty principle the 
$\bar{n}$-nucleus cross sections 
should be almost $A$-independent, apart for fluctuations due 
to nuclear surface effects. Consequently the $\bar{p}$-nucleus 
cross sections should depend on the target because of its 
electric charge only. The underlying argument is that 
most of the existing models (see the suggested references 
\cite{dov1,amsl,kerb,walch,wei93,eri88}) 
and analyses\cite{brue91,aar,bonn1} establish that the 
annihilation 
process takes place when the centers of mass of the 
antinucleon and of the target nucleus 
are at a relative distance $d$ such that 
$R_{nucleus}$ $<$ $d$ $<$ $R_{nucleus}+\Delta$, where 
$\Delta$ $\sim$ 1 fm (or smaller, depending on 
the model) does not depend too much on the target. 
So the annihilation is equivalent to 
a measurement of the projectile-target relative distance 
with uncertainty $\Delta$ $<$ 1 fm, and this measurement is 
incompatible with a relative momentum $<<$ 200 MeV/c. 

To see it another way, we distinguish between two classes 
of nuclear reactions. On one side, inelastic reactions 
where the entire nucleus is involved, as in compound nucleus 
reactions, but the underlying projectile-nucleon 
processes are elastic (e.g. neutron induced nuclear 
reactions). In this case the characteristic reaction 
region coincides approximately with the target nucleus. 
Then the uncertainty $\Delta$ coincides approximately 
with the nuclear radius. On the other side, we find 
reactions where a strong inelasticity is present at the 
projectile-nucleon level. In this case reactions deep  
inside the nuclear volume are rare, the reaction region 
is a shell at the surface of the target nucleus, with 
thickness $\Delta$, and $\Delta$ is approximately the 
same for all the possible targets. 

The consequence of the limitations imposed by the 
uncertainty principle is that for antinucleon 
momenta $k$ $<<$ $1/\Delta$ the total reaction cross section 
becomes much smaller than its possible unitarity limit. 
This is also established by the well known\cite{ll1} 
low energy limit for the phase shifts: 
$\delta_l$ $\propto$ $k^{2l+1}$ for $k$ $\rightarrow$ 0. 
The unitarity limit is reached when a 
partial wave is completely absorbed in the reaction 
process, which means $exp(i\delta_l)$ $=$ 0, i.e. 
$Im(\delta_l)$ $=$ $\infty$, so the unitarity limit 
cannot be attained at small enough $k$. Uncertainty 
considerations suggest that for $k$ $>>$ $1/\Delta$ 
it is possible, for strong enough reactions, to 
saturate the unitarity limit, while for $k$ $<<$ 
$1/\Delta$ we are in the situation where $\delta_l$ 
$=$ $O(k^{2l+1})$, whatever the strength of the reaction. 
A paradoxical consequence is that 
a smaller $\Delta$ corresponds to what would be a 
stronger reaction at large energies, so that at low 
energies ``stronger'' interactions can lead to a smaller 
reaction rate. This fact can be verified in  
optical potential treatments. 

On the ground that the projectile wavefunction $\Psi$ 
is completely damped within a range $\Delta$ (i.e 
$\vert \Psi\vert$  
is large for $r$ $>$ $R_{nucleus}+\Delta$ and very small 
for $r$ $<$ $R_{nucleus}$) 
it is straightforward to demonstrate that for the 
scattering length $\alpha$ we have (approximately):

$Im(\alpha)$ $\approx$ $-\Delta$, 

$Re(\alpha)$ $\approx$ $+R_{nucleus}$. 

\noindent
Indeed, the $\Psi$ damping 
requirement implies 
for the logarithmic derivative $\vert \Psi'/\Psi\vert$ $\approx$
$1/\Delta$. This is an obvious geometrical fact, but 
in more physical terms it is a consequence of the 
uncertainty principle. 
Together with the matching condition between the logarithmic 
derivatives of the free motion wavefunction and of 
the wavefunction in the annihilation region  
$\vert \Psi'/\Psi\vert_{r=R_{nucleus}+\Delta}$ $=$ 
$\vert k\cdot cotg\{k(R_{nucleus}+\Delta-\alpha)\}\vert$ 
this leads to the previous $\alpha$-values in the limit 
$k$ $\rightarrow$ 0. 

These values of course are deduced from approximate 
equations, so they represent just estimates, however 
they suggest that the antineutron 
annihilation cross sections should  
not show a $systematic$ increase with the target mass 
number $A$. Such an increase could be present for 
antiproton annihilations, but because of Coulomb effects 
only. When going to any specific target nucleus, 
non-systematic effects could be present, especially 
related with the structure of the nuclear surface. 
An example is given in ref.\cite{glw174}. There, anomalous 
behaviors are related with the non-sphericity of the nucleus. 
Another exception should be represented by neutron-halo 
nuclei, because in this case the 
annihilation range could be much larger than 1 fm. 
Also the composition of the nuclear surface in 
terms of protons or neutrons could be important, since 
all models attribute a strong isospin dependence to the 
antinucleon-nucleon interaction. 

The exposed mechanism has an interesting consequence in 
the case of optical potential analyses: an increase of the 
strength of the imaginary part of the optical potential can 
lead to a decrease of the consequent reaction rate at 
small momenta\cite{pro1,pro3,n1,n2,n3}. In the above 
language, an increase 
in the potential strength leads to a decrease in the 
size parameter $\Delta$, since the absorption of the 
projectile wavefunction takes place in a shorter range. 
Also modification of other parameters 
(radius, diffuseness, etc) leads to consequences that 
are not necessarily the most obvious ones. An example is 
given in the next section. 

\section{Comparison between $\bar{np}$ and $\bar{p}p$ total 
annihilation cross sections.} 

In fig. 1 we show the TNA and TPA on Hydrogen, together with 
two fits by energy-independent optical potentials.  
For the $\bar{p}p$ case the total interaction includes 
the electrostatic potential of a spherical charge distribution 
with radius 1.25 fm. This charge radius is $\sqrt{2}$ times the 
charge radius of the proton, to take into account both the 
proton and the antiproton extended charges.  
Details on the potential are given below. 
A third optical potential curve shows what 
would be the $\bar{p}p$ cross section in absence of electrostatic 
interactions, within the same optical model. 

\begin{figure}[htp]
\begin{center}
\mbox{
\epsfig{file=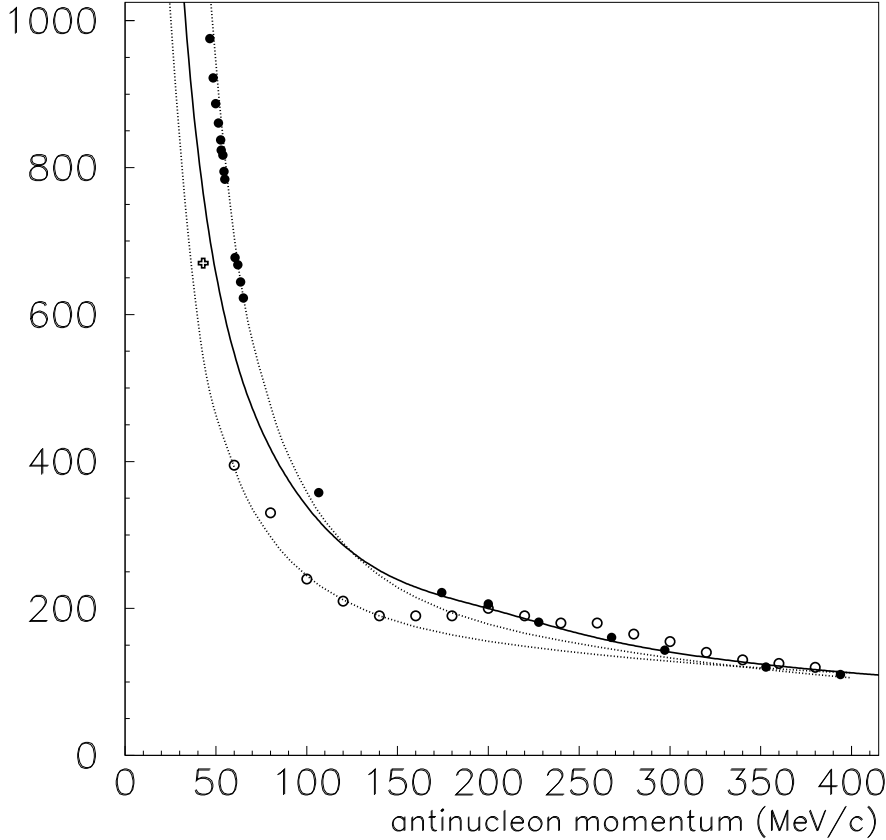,width=0.9\linewidth}}
\end{center}
\caption[]
{\small\it \label{fig1}
Antineutron (empty circles) and antiproton (full circles) 
total annihilation cross sections (mb)
measured by the Obelix experiment\cite{feli,obe1,obe3}
(called TNA and TPA, respectively, in the text).  
The empty crosses reproduce the two low-energy 
$\bar{n}p$ total annihilation points measured in \cite{mutch}.  
Error bars are not reported. 
The two dotted lines 
correspond to optical potential fits (see text for details). 
The solid line represents the $\bar{p}p$ annihilation 
cross section after Coulomb effects have been subtracted, as 
described in the text. 
The lower energy part of this curve has been calculated 
by extrapolating the optical potential fit of the $\bar{p}p$ 
data and by removing the electrostatic part 
of the potential. For $k$ $>$ 30 MeV/c the Coulomb effects have 
been subtracted from the actual $\bar{p}p$ points, not from 
the potential fit (for this 
reason the solid curve is larger than the dotted curve 
for $k$ $>$ 130 MeV/c). }
\end{figure} 

What we notice first is that TNA and TPA are reasonably similar 
for $k$ $>$ 200 MeV/c, but below 200 MeV/c the TNA falls clearly 
below the TPA, and this fall can not be justified in terms of 
Coulomb interactions only, although these contribute. We could 
speak of ``low energy fall'' with respect to a ``background''  
which is similar to the $\bar{p}p$ case. 
Alternatively one can think 
that the TNA sums two contributions: a monotonous 
background that is proportional to the $\bar{p}p$ 
annihilation rate (Coulomb effects apart) but 
lower,  
and a broad peak (resonance?) at $k$ $\sim$ 
200$-$300 MeV/c. 
This broad peak would be hidden by the isospin-0 
background and by the low energy Coulomb rise in the TPA 
case. 
Some of the available data at momenta $>$ 400 MeV/c 
(see \cite{bende} for a more systematic review of 
these data) could support the ``peak'' idea, since at these 
momenta the $\bar{n}p$ annihilation rate seems to be lower 
than the $\bar{p}p$ one, but it is not unequivocally  
clear how much lower. Indeed, one $\bar{n}p$ point 
at 700 MeV/c by \cite{bane85} is clearly below the 
$\bar{p}p$ annihilation, while 
data reported in \cite{arms87} 
are not very different from the corresponding $\bar{p}p$ 
data\cite{brue90} in the region 300-500 MeV/c. 
On the other side, 
some data at very low energy\cite{mutch} support 
the ``gap'' idea, since they show a rise of the 
$\bar{n}p$ annihilation rate $\beta\sigma$ 
below 50 MeV/c. These data 
have been reported in the figure, and the one at 20 MeV/c 
lies on the optical potential extrapolation of the $\bar{p}p$ 
data, once the electrostatic part of the potential has 
been removed. The authors of ref.\cite{mutch}  
also report some potential 
predictions\cite{dr1,paris} that support the existence 
of a minimum for the $\bar{n}p$ annihilation rate 
$\beta\sigma_{ann}$ at 50-100 MeV/c, and 
a set of scattering length 
predictions\cite{dr1,paris,bp} that show no great differences 
between $\bar{p}p$ and $\bar{n}p$ expected scattering lengths.  
They also report a collection of pre-Obelix $\bar{n}$ data. 

We must remark that the normalization 
of antineutron fluxes is a delicate matter, with the consequence 
that both a comparison of antineutron data coming from different 
experiments, and a comparison of antineutron and antiproton 
data, must be taken with the greatest caution. So we will 
generically speak of a ``gap/peak structure'' of the $\bar{n}p$ 
annihilation rate compared to the $\bar{p}p$ one. We will 
show in the next section that a comparison with nuclear data 
supports the existence of this gap/peak structure. 
Whatever the interpretation, there are some important 
remarks: 

1) The strong oscillation of the TNA with respect to the 
TPA (see also fig.4) is an evidence of the fact that 
different physical mechanisms are dominating in the two 
cases in the considered momentum range. 
If a resonance is present this is obvious. If there is 
no resonance, 
according to the partial wave analyses presented in 
refs.\cite{feli,maha} 
the TNA P-wave has its maximum at $k$ $\approx$ 250 MeV/c 
where it attains the value 120-140 mb. 
Our optical potential fit (which seems good for the TPA) 
fixes the TPA P-wave maximum to about 80 mb at 150 MeV/c. 
Semiclassical intuition fixes the corresponding impact 
parameters $b$ $=$ $L/k$ 
to 1.25 fm (TPA) and 0.8 fm (TNA). Then, purely 
geometrical considerations suggest a ratio, between the 
size of the two P-wave outcomes at their peak momentum, 
TNA/TPA $\sim$ $(0.8/1.25)^2$ $\approx$ 0.4 instead of 
the found value 1.6. The most obvious conclusion would be 
that the P-wave interaction is much more effective in 
the $\bar{n}p$ case, where it acts at smaller impact 
parameters (by a factor 0.6). This  
short range interaction does not affect the S-wave 
contribution, which is clearly larger in the $\bar{p}p$ 
case (even after subtracting Coulomb effects) at 
momenta over 30 MeV/c. At lower momenta 
(that means longer range) the two S-wave 
contributions could be equal, according 
to\cite{mutch} and references therein. So the  
geometry and strength of the isospin-1 and isospin-0 
channel interactions must be different, as far as these 
semiclassical considerations can be trusted. 

2) Absurd as it may sound, the fact that at small momenta 
TNA are sensibly smaller than TPA could mean that 
the interactions are stronger in the $\bar{n}p$ case than 
in the $\bar{p}p$ one. This can be immediately seen in the 
choice of the optical model parameters. We used a Woods-Saxon 
form, with all parameters, but one, equal for TPA and TNA. 
TPA: imaginary strength 8000 MeV, real strength 46 MeV 
(attractive), imaginary radius 0.52 fm, real radius 1.89 fm, 
real and imaginary diffuseness 0.15 fm. These values 
fall in the ranges used by previous authors 
to fit elastic data\cite{brue86}, and its imaginary 
part is pretty similar to other previously used 
optical potentials (e.g. the one of ref.\cite{kw86}). 
Many other sets of parameters, and also different potential 
shapes, can lead to similar results for 
the available low-energy data. 
Starting from this peculiar set, 
to obtain the TNA curve it is sufficient to increase 
the imaginary radius to 0.75 fm. Alternatively, one can 
obtain the TNA curve by leaving the imaginary radius at 
0.51 fm and increasing the imaginary strength to about 
12000-16000 MeV. So, to get a smaller cross section we 
need a ``stronger'' potential.  
This ``inversion'' behavior is another manifestation of 
the general shadowing mechanism discussed in 
section 3. In an optical potential model at small $k$ 
an increase, e.g., of the imaginary strength $W$ produces an 
increase of the reaction cross section for small values 
of $W$ only. After a certain threshold,  
further increases of $W$ lead to a decrease of the 
reaction cross section\cite{pro1,n1,n2,n3,pro3}. 
At a qualitative 
level this can be explained by the uncertainty principle: 
an increase in the strength of the imaginary part of the 
potential decreases the thickness of that spherical shell, 
surrounding the proton/nucleus target, where annihilations 
are supposed to take place. This introduces larger 
gradients in the projectile wavefunction. These larger 
gradients imply a larger logarithmic derivative, and therefore  
a smaller (imaginary part of the) scattering length. 

3) In the case of $\bar{p}p$ annihilations below 600 MeV/c 
many quite different choices of potential parameters 
can fit them\cite{n1,n2,n3} (including a 
pure imaginary potential) without introducing a 
dependence of the parameters on energy. 
All these fits share some common features. They 
reproduce the data very well below 150 MeV/c and above 300 MeV/c. 
The fits remain satisfactory up to 
600 MeV/c.  In the region 150-300 MeV/c the fits 
stay slightly below the data. Alternatively, it is possible 
to choose the parameters so as to reproduce the data from 
150 to 600 MeV/c, while at lower momenta the fits are above 
the data. So, we could say that an energy independent 
optical potential can produce a ``background'' around which 
a slight oscillation is present (positive for $k$ 
in the range 150-300 MeV/c, or negative for 
$k$ $<$ 150 MeV/c). In the TNA case the situation is 
qualitatively the same, but the oscillation is much 
stronger and more evident. So it seems that an 
energy-independent optical potential fit can be 
well related 
with the isospin-0 channel, which seemingly consists 
of the ``background'' only. A background 
with the same shape is also present 
in the isospin-1 channel, but this channel also 
contains an oscillation that is not reproduced 
by our set of optical potentials. As already remarked, 
we have no means to establish the 
nature of this oscillation. 

4) We have also tried to change our optical potential shape 
so as to produce resonances that fit 
the gap/peak structure which is seen in the $\bar{n}p$ 
annihilation. We can not exclude that more attempts can 
lead to a good reproduction of the data in fig.1 
(in the described ``inversion plus resonance'' regime 
there is no predictable relation between the changes of 
the potential and the shape of the output, which makes such 
attempts rather frustrating). We have not been able to 
reproduce 
satisfactorily these data, and perhaps more sophisticated 
potentials are necessary. It was however easy to get 
shapes that $qualitatively$  
looked similar. We tried with two kinds of potential: 
(i) Woods-Saxon (with attractive elastic part) 
(ii) Woods-Saxon like the previous one, 
plus a repulsive elastic surface barreer of gaussian 
shape (of the kind $exp[-(r-r_o)^2]$). In both cases  
to get a broad peak where it must be it was necessary to 
enlarge the range of the elastic part to 2.5-3 fm. This 
is probably due to the fact that, despite the smallness 
of the radius of the imaginary part, most of the 
annihilations take place at $r$ $\approx$ 1 fm. So only 
a much wider attractive well can produce metastable 
states, within such models. Perhaps in a more 
sophisticated potential model, where 
different spin channels are clearly separated, it 
is possible to get a resonance in a spin channel where 
annihilation is less effective, so without the need for 
such a long range for the potential. 
We think that these 
interesting data really demand for a higher level potential 
analysis. 

\section{Annihilation on nuclei. Shadowing in the charge 
ratio.} 

Because of the shadowing, a complete analysis of 
annihilations in nuclei below 60 MeV/c is pretty complicated. 
A simple comparison of different sets of data reveals 
interesting features, however, and we 
can also hope to understand something 
by a PWIA analysis, especially at momenta over 100 MeV/c. 
In particular, the energy dependences 
of the $\bar{n}p$ annihilation cross 
section, of the $\bar{n}$-nucleus one, and 
of the ratio of $\bar{p}n$ to the $\bar{p}p$ annihilation in 
nuclei, show evident correlations. 
Of course nuclear shadowing effects are completely 
absent in PWIA, and this must be taken into account, 
taking with great care the calculations below 100 MeV/c.
At larger momenta the main missed effect is the eclipse 
effect\cite{glau} which implies 
an almost energy independent surface absorption 
factor $\approx$ 0.5. We take this factor into account by 
renormalizing our PWIA predictions with one datum 
at a specified momentum value over 200 MeV/c. 

At small momenta however (below $\sim$ 200 MeV/c) 
shadowing is something more complicated than a mere 
eclipse effect. The eclipse effect 
could never bring $\bar{p}$ annihilation rates to  
decrease with increasing $A$. So we 
may only $define$ as ``shadowing'' the departures 
from PWIA predictions, and remark that they are consistent. 
As we show later, anyway, 
a certain part of the smallness of the low energy 
$\bar{p}$-Deuteron annihilation rate with respect to the 
$\bar{p}p$ one is not a shadowing effect, but just due 
to the smallness of the $\bar{p}n$ annihilation rate 
below 150 MeV/c. 

In fig.2 we show the $\bar{n}$-nucleus data, relative to 
momenta between 180 and 280 MeV/c, for several nuclear species 
from $^{12}$C to $^{207}$Pb (see \cite{ableev} for more 
details). 
These data are consistent with an $A^{2/3}$ law, expressing 
dominant surface absorption, $among$ $them$. However 
the $A^{2/3}$ law found in these data 
cannot be generalized to $\bar{n}p$ or 
$\bar{p}p$ (Coulomb subtracted) data in the laboratory 
frame, while this generalization is 
only approximately possible in the center of mass frame. 
At 180 MeV/c, the $^{12}$C or the $^{207}$ Pb TNA, 
divided by $A^{2/3}$, give about 150 mb. The 
corresponding $\bar{n}p$ and (Coulomb subtracted) 
$\bar{p}p$ cross sections are $\approx$ 200 mb 
at the same laboratory momentum. At the same center of 
mass momentum, which corresponds to 360 MeV/c in the 
laboratory, the $\bar{n}p$ total annihilation cross section 
is about 130 mb, and the $\bar{p}p$ one is slightly 
smaller than the $\bar{n}p$ one. 
So, in both cases 
we find deviations from the $A^{2/3}$ law at 
$k_{cm}$ $\approx$ 100 MeV/c, but these deviations 
are of opposite sign if the data are compared at the 
same laboratory or center of mass momentum. 
It is difficult to establish whether it is more proper 
to compare data relative to different nuclear species 
at the same laboratory or center of mass 
momentum. Impulse approximation based models suggest the same 
laboratory momentum, compound nucleus inspired models 
prefer the 
opposite choice. Indeed, in the former case direct momentum 
exchange is between the projectile and one of the target 
nucleons only, whose average initial momentum is zero 
in the laboratory. In the latter, the projectile momentum is 
given directly to the entire target nucleus. 

\begin{figure}[htp]
\begin{center}
\mbox{
\epsfig{file=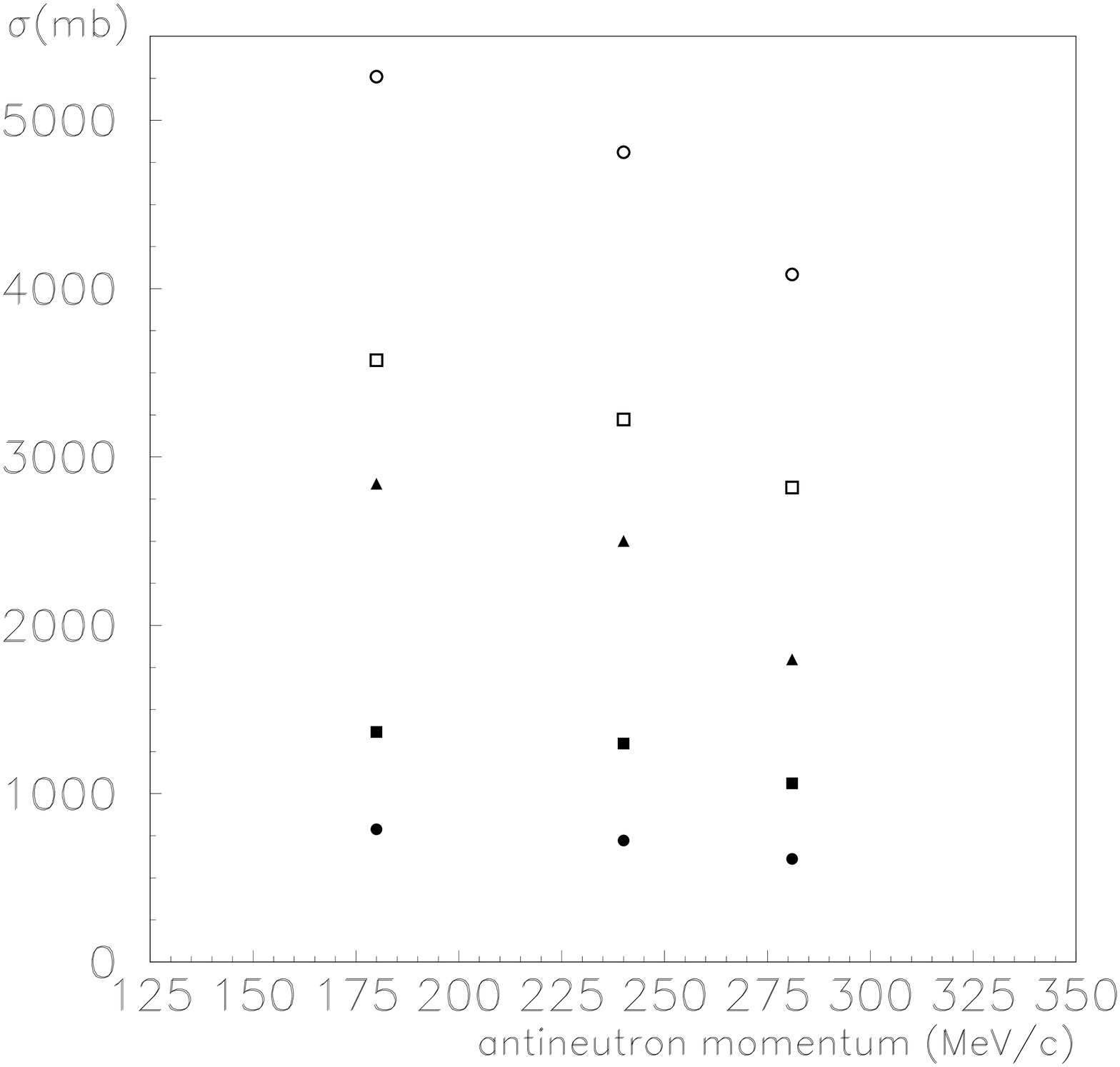,width=0.9\linewidth}}
\end{center}
\caption[]
{\small\it \label{fig2}
$\bar{n}$-nucleus total annihilation cross sections for 
C (full circles), Al (full squares), Cu (full triangles), 
Sn (empty squares), Pb (empty circles). All the data have 
been taken at antineutron momenta  
180, 240 and 281 MeV/c in the laboratory 
(see \cite{ableev} for more details). 
}
\end{figure} 
\begin{figure}[htp]

\begin{center}
\mbox{
\epsfig{file=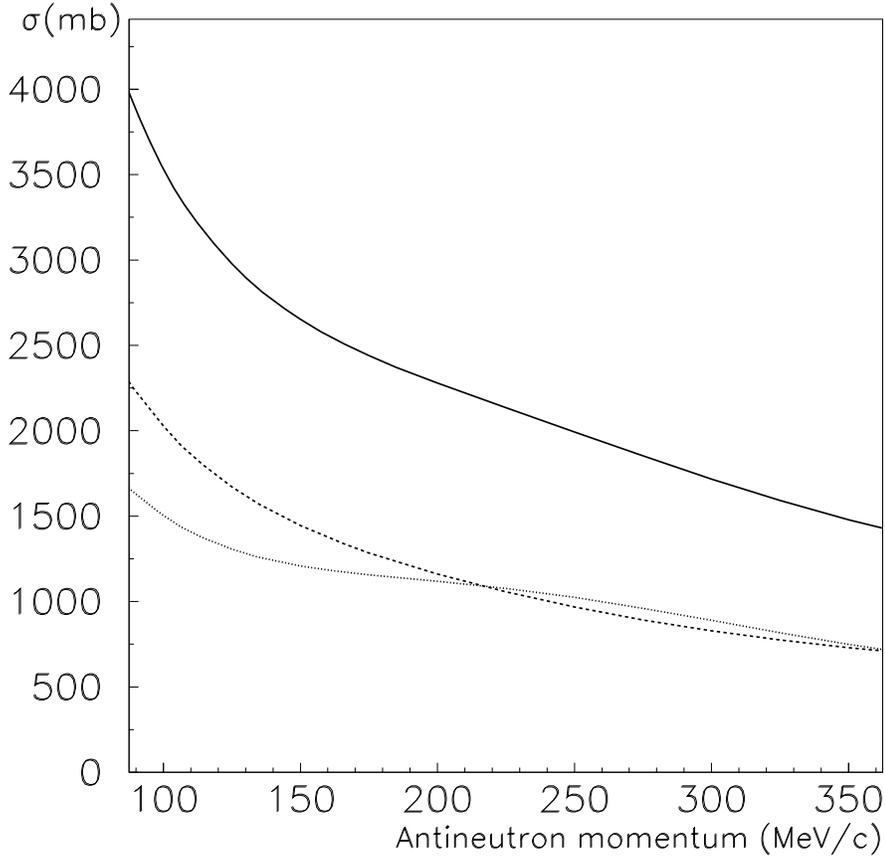,width=0.9\linewidth}}
\end{center}
\caption[]
{\small\it \label{fig3}
PWIA calculation of the annihilation cross section 
$\bar{n}-^{40}$Ca. 
The reported curves have been multiplied by 0.5 to take  
absorption into account. This (approximate) factor has been 
estimated from the data in fig.2. Continuous curve: the total 
annihilation cross section. Dashed line: contribution from the 
$\bar{n}n$ annihilations. Dotted line: contribution from 
the $\bar{n}p$ annihilations. 
}
\end{figure} 
\begin{figure}[htp]
\begin{center}
\mbox{
\epsfig{file=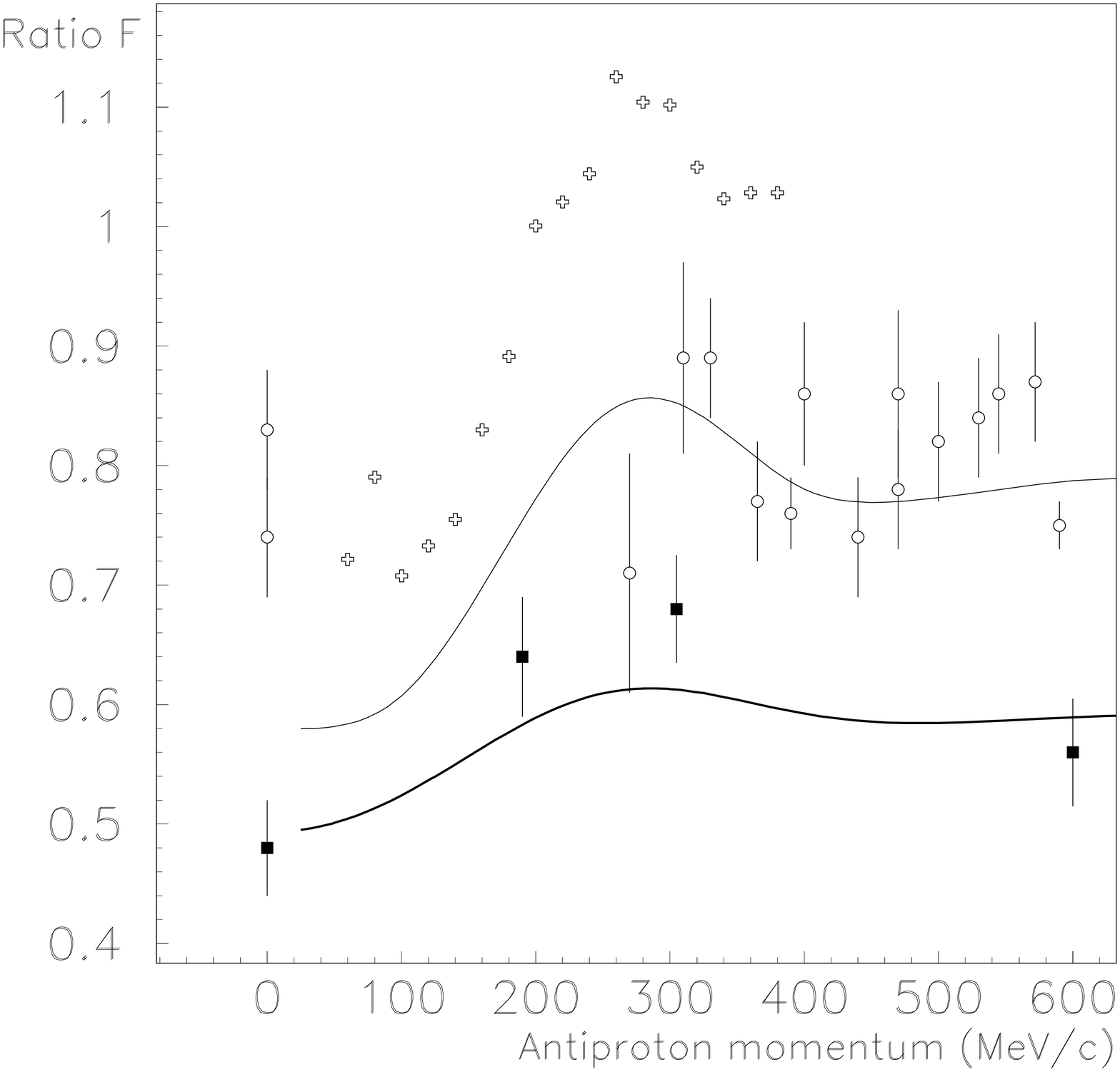,width=0.9\linewidth}}
\end{center}
\caption[]
{\small\it \label{fig4}
Ratios $F_A$ of antiproton annihilations on 
neutrons to antiproton annihilations on protons 
in a nucleus, for $A$ $=$ 2 (Deuteron - empty circles) 
and 4 ($^4$He - full squares). For more details about 
these data see \cite{nb2}. The empty crosses represent, 
for comparison, the ratio $F_1$ between the $\bar{n}p$ 
total annihilation cross sections reported in 
\cite{feli} and the ``Coulomb subtracted'' 
$\bar{p}p$ total annihilation cross sections (continuous 
line in fig.1). The two continuous lines represent PWIA 
fits of $F_2$ and $F_4$. To produce these fits the 
$\bar{n}p$ annihilation rates have been rescaled by an 
energy-independent factor $\alpha_A$. For Deuteron 
$\alpha_2$ $=$ 0.8, and for $^4$He $\alpha_4$ $=$ 0.6. 
More details on these PWIA fits are in the text. 
}
\end{figure} 

An analogous,  
systematic comparison of all the available $\bar{p}$-nucleus 
data has been presented elsewhere\cite{ne991}. 

Very recently a new series of $\bar{n}$-nucleus data 
down to 70 MeV/c 
has been presented\cite{brx}, which show a respected 
$A^{2/3}$ law between different nuclear species. Since these 
data have been presented as ``preliminary'', we can't 
discuss them in detail. 

Considering the energy behavior of the TNA on nuclei shown in 
fig.2, the most interesting feature is the 
systematic presence of a shoulder. However this phenomenon 
seems less evident in the just quoted preliminary data. 
TPA data in this region are not enough to to give us 
shape information. However, the optical potential calculations 
that we have reported in \cite{n2} 
suggest that the Coulomb enhancement at low energies can 
easily hide such details in the energy dependence of 
the annihilation data. 

Different relevant mechanisms must be taken into account 
in the momentum range 100-300 MeV/c: 

1) The tail of the low energy shadowing phenomenon.  
Presently we have no knowledge of TPA and TNA 
on heavy nuclei at $k$ $<<$ 100 MeV/c, so we do not know 
whether the discussed shadowing is a systematic phenomenon 
(as we suggest in the previous section 2) 
or is simply peculiar of some light nucleus. And in light 
nuclei we do not have enough TPA data in the region 100-300 
MeV/c to understand completely the transition between 
large and low energy regimes. There are several sets of data, 
but regretfully we do not have a satisfactorily continuous 
set of 
data relative to the same projectile on the same target 
(with the exception of the $\bar{p}p$ and $\bar{n}p$ cases). 

3) The nuclear Fermi motion. The Fermi momentum, which 
represents a rough cutoff of the nuclear momentum distribution, 
is $\approx$ 200 MeV/c. This means that for $k$ $<$ 200 MeV/c 
the statistical distribution of the relative $\bar{n}$-nucleon 
motion contains the zero-energy point, and that it is 
easy for a target nucleon to be faster than the projectile. 
So the projectile momentum 200 MeV/c represents a kind of 
special borderline in the physics of annihilations on nuclei. 

4) The $\bar{n}p$ annihilation cross section presents 
the above discussed gap/peak structure below 
300 MeV/c. The nuclear Fermi motion will soften this 
structure, but perhaps not enough to make it completely 
disappear. 

A PWIA calculation can give some qualitative 
understanding about the 
roles of the gap/peak structure and of the Fermi motion in 
the nuclear TNA, and also about the relative role of 
neutrons and protons in the target. 

In fig.3 we present a PWIA calculation of the TNA on $^{40}$Ca. 
In fig.4 we show a PWIA calculation of the $F_A$ ratio on 
Deuteron and $^4$He. We recall that we have defined $F_A$ 
as the ratio between $\bar{p}$ annihilations on neutrons 
and protons inside a nucleus with mass number $A$. 
In figs. 5 and 6 we show PWIA calculations of $\bar{p}$ 
annihilations on deuteron and $^4$He. 

As a starting point we have used an equation contained in 
$\cite{n1}$. Although this equation has been justified 
there by a long series of mathematical passages, its meaning 
is quite simple and can stand without demonstration: 
at PWIA level, any event rate on a nuclear target, leading to 
a well defined final state, is an incoherent sum of the 
rates of all the possible $\bar{n}n$ and $\bar{n}p$ events 
leading to the same final state, weighted by the 
nuclear momentum distribution of the target nucleon. 
By ``rate'' we mean a cross section divided by the 
corresponding flux of colliding particles. 
Under the simplifying assumption 
that different final states sum incoherently 
we simply have: 

\begin{equation}
\beta(k,0)\sigma(k)\ =\ 
\int N(\vec k')\beta(\vec k,\vec k')
\sigma(\vert\vec k-\vec k'\vert) d^3k',
\label{pwia}
\end{equation}
where $\beta(\vec k,\vec k')$ is the relative velocity between 
two colliding particles with momenta $\vec k$ and $\vec k'$. 

This nuclear PWIA average 
of the antinucleon-nucleon processes 
clearly misses any eclipse or shadowing effect 
(which, e.g., produces a TNA proportional to $A^{2/3}$ 
at large energies). So, the rise of the PWIA 
curve in fig.3 below 150 MeV/c is unreliable (it is just 
proportional to the average rise of antiproton and 
antineutron annihilation cross sections on a proton). 
At momenta larger than 150 MeV/c the PWIA should be more 
reliable, apart for a slowly energy dependent  
eclipse factor, and could clarify the 
effects of the Fermi motion and the 
different role of protons and neutrons in the target. 

For the $^{40}$Ca shells we have used harmonic oscillator 
eigenfunctions. Some attempts with different kinds of 
states have shown no qualitative differences, unless 
the nuclear size is changed to unphysical values. In 
addition, the most external wavefunctions do not show 
great differences (in the energy dependence of the outcome) 
with respect to the 
S-wave ground state. This is important, because in a 
more proper DWIA (distorted wave impulse approximation) 
treatment the internal shells would 
be scarcely involved in the annihilation process. 
For Deuteron and $^4$He (in fig.4) 
experimental single particle 
momentum distributions\cite{fm} have been used. Again, 
within this simple model we did not find a great 
sensitivity of the results on the peculiarities of the 
momentum distribution, taking into account that 
many details that could be present in the numerator 
and in the denominator are lost in the ratios 
of fig.4. With Deuteron, we surely missed 
the angular dependence of the large momentum (D-wave 
dominated) part of the distribution. 

To calculate the PWIA curves of figure 3 we need 
$\bar{n}p$ and $\bar{n}n$ 
annihilation rates. For fig.4 we need $\bar{p}p$ and $\bar{p}n$
annihilation rates, both subject to the effect of the 
Deuteron or $^4$He charge. The $\bar{n}n$ annihilation rate 
has been assumed to be equal to the Coulomb subtracted 
$\bar{p}p$ one (i.e. the continuous curve of fig.1). 
In fig.4 we use the same cross sections, so we neglect  
the effects of the nuclear electrostatic attraction. On 
the basis of the discussion in section 2 we 
assume that they do not affect the $F_A$ ratio, 
although they surely enhance both the 
$\bar{p}p_{in\ nucleus}$ and the $\bar{p}n_{in\ nucleus}$ 
rates. Actually a look at the data in fig.4 shows that 
(with large error bars) 
$F_4/F_2$ $\simeq$ 0.48/0.78 $\simeq$ 0.61 at zero energy, 
and 0.54/0.82 $\simeq$ 0.66 at about 600 MeV/c. In 
the case of a different action of the nuclear 
electrostatic potential on the $\bar{p}p$ and $\bar{p}n$ 
cross sections, this effect would be practically absent at 
600 MeV/c and would depress the $F_4$ ratio with respect to 
the $F_2$ one at momenta $<<$ 100 MeV/c. The large 
error bars and data fluctuations do not allow for easy 
conclusions, however there are no special reasons to assume 
a different Coulomb effect on $\bar{p}p$ and $\bar{p}n$ 
reactions, if both the proton and the neutron are bound in 
the same nucleus. 

The Obelix $\bar{n}p$ data cover the range 60-400 MeV/c, 
and the $\bar{p}p$ ones start at 30 MeV/c. We therefore had to 
assume values for these cross sections for momenta outside these 
ranges. With $\bar{p}p$ data at momenta over 200 MeV/c we 
have relied on the data in ref.\cite{brue90}, and at momenta 
below 30 MeV/c on our optical potential fit (after subtraction of 
Coulomb forces as in fig.1). For the $\bar{n}p$ case,  
in the region $k$ $>$ 400 MeV/c the 
$\bar{n}p$ data have been assumed to be equal to the 
$\bar{p}p$ ones, as suggested 
by the measurements in ref.\cite{arms87} and \cite{brue90} 
for momenta over 300 MeV/c 
(see e.g. \cite{bende}, 
fig.9, for a comparison of the $\bar{n}p$ and $\bar{p}p$ 
data in that momentum range). A look at fig.4 does not 
help too much. Indeed, we may say that 
the nuclear Fermi motion approximately 
averages the ``free'' annihilation 
rates within a 200 MeV/c range. Then 
the point of view of equal ``free'' $\bar{p}p$ and $\bar{p}n$ 
annihilation rates 
at momenta in the range 300-600 MeV/c 
can be supported by the absence of a clear decreasing trend 
in the set of Deuteron points over 300 MeV/c. On the contrary, 
the Helium point at 600 MeV/c supports the possibility  
of a decrease of the free $\bar{p}n/\bar{p}p$ 
ratio, as suggested by the measurement 
in ref.\cite{bane85}. Anyway, the scatter of data in the 
deuteron case, and the lack of data in the $^4$He case 
prevent us from precise conclusions. 
In the limit of zero 
energy, we have adopted the low energy parametrization 
of the $\bar{n}p$ cross section contained in 
references\cite{feli,maha}. It implies a decrease 
of the $F_A$ ratios at very low energies, which is 
more suitable for fitting the zero-energy $^4$He point. 
Adopting the low energy parameters suggested by 
refs.\cite{mutch,bp,dr1,paris} would on the contrary rise 
the zero energy $F_A$ value, which is more coherent with 
the two zero-energy Deuteron points. 

The upper curve in fig.3 can be taken as a qualitative 
confirmation of the effect of the $\bar{n}p$ gap/peak structure 
in producing a very soft shoulder in the $\bar{n}-$nucleus 
annihilations at 150-300 MeV/c
In fig.3 we also show the separate contributions of the 
$\bar{n}p$ and $\bar{n}n$ reactions. 
Evidently the $\bar{n}n$ reaction alone would not interrupt 
the trend of a positive second derivative that characterizes 
expectance and data at both lower and larger momenta. 
Within PWIA the shoulder is a direct consequence 
of the gap/peak structure of the $\bar{n}p$ cross section. 
The Fermi motion, as predictable, softens this gap/peak 
structure, but not completely. We must also observe that 
the produced shoulder is much less pronounced than most of 
the ones seen in fig.2. Our attempts with 
different shapes for the nuclear shells show that the 
shoulder evidence is enhanced by a larger radius of each 
nuclear shell (i.e. a narrower momentum distribution), 
however one must arrive at slightly unrealistic 
nuclear dimensions in order to obtain more evident 
shoulders. 

The strength of the rise of the PWIA cross sections 
below 100 MeV/c 
is probably unphysical, since we know that shadowing 
could be very active in that region (although the quoted 
recent preliminary data \cite{brx} confine 
shadowing to $k$ $<$ 70 MeV/c and with good 
approximation 
reproduce the predictions of fig.3). In ref.\cite{n2} 
some optical potential shapes possibly corresponding 
to low energy $\bar{n}$-nucleus annihilations are reported, 
and they are almost energy-independent in the region 
100-300 MeV/c. These optical potentials were based on 
Coulomb-subtracted $\bar{p}p$ data and on nuclear cross 
sections at momenta over 200 MeV/c and so 
have no relation with that gap/peak structure 
that is peculiar of the $\bar{n}p$ interaction. 
The data reported in \cite{brx} confirm the IA behavior 
reported in fig.3 down to 70 MeV/c, rather than the 
these optical potential predictions. 

The systematic presence of the gap/peak structure 
in the TNA/TPA ratio is much more 
evident in fig.4, where we report the $F_A$ ratio for 
$A$ $=$ 1 ($\bar{n}p$ data coming from the Obelix 
experiment\cite{feli} divided by the Coulomb subtracted 
$\bar{p}p$ annihilation cross section represented 
by the continuous line in fig.1), for $A$ $=$ 2 
(Deuteron) and $A$ $=$ 4 ($^4$He). These Deuteron and 
$^4$He data come from several experiments 
and are reviewed in ref.\cite{nb2}. 

Looking at fig.4 one immediately notices  
that all deuteron points lie above any of the $^4$He points, 
and the proton target data are, in the average, clearly 
over the nuclear target data (we remark that, 
although the proton target ratios have been calculated 
using the ``Coulomb subtracted'' $\bar{p}p$ annihilation 
cross section, the effect of the Coulomb subtraction is 
negligible above 100 MeV/c). 
The second observation concerning the data reported in fig.4 is 
that a certain degree of correlation 
between the shapes of the three groups of data is present. 

Since $Z/A$ $=$ 1 both in Deuteron and 
$^4$He, and the average value of the free $F_1$ ratio 
is about 1, clearly 
PWIA cannot reproduce the average value of 
these nuclear $F_A$ distributions. 
For producing the fits contained in fig.3 we have multiplied 
the $\bar{n}p$ total annihilation cross section by a constant 
factor $\alpha$, which is 0.8 for Deuteron target, and 0.6 
for $^4$He target. 
At the largest energies of the considered 
range the PWIA ratio should be reliable. 
The known eclipse effect should more or 
less cancel between numerator and denominator. Anomalous 
shadowing of the kind discussed in section 3 is not strong 
over 100 MeV/c. So the 
reason for $\alpha_{^4He}$ $<$ $\alpha_D$ $<$ $\alpha_p$ 
$\equiv$ 1 is unexplained. A possibility is final state 
rescattering. This should not be very effective in 
total annihilation cross sections, 
but in the $F_A$ case it can modify the number of charged 
particles that is used to decide whether the annihilation 
was on a proton or on a neutron. It is not unlikely that 
this $A$-dependent renormalization effect 
has this origin, however one should 
analyze carefully how each experimental point has been 
measured. 

In the case of $^4$He the PWIA fit reproduces 
satisfactorily the 
behavior of the data. We may say that the gap/peak 
structure of the free $\bar{n}p/\bar{p}p$ ratio is 
reproduced by the $F_4$ ratio. 
In the Deuteron case at zero-energy the fit is not 
satisfactory, while over 200 MeV/c some correlation 
seems to be present. Somehow the presence of the PWIA 
peak at 280 MeV/c (which is a consequence of the 
gap/peak structure of the free ratio) is reflected in 
the presence of a maximum in the data at 300 MeV/c,  
but the large error 
bars and the spread of the data do not allow for 
more quantitative conclusions. 

\begin{figure}[htp]
\begin{center}
\mbox{
\epsfig{file=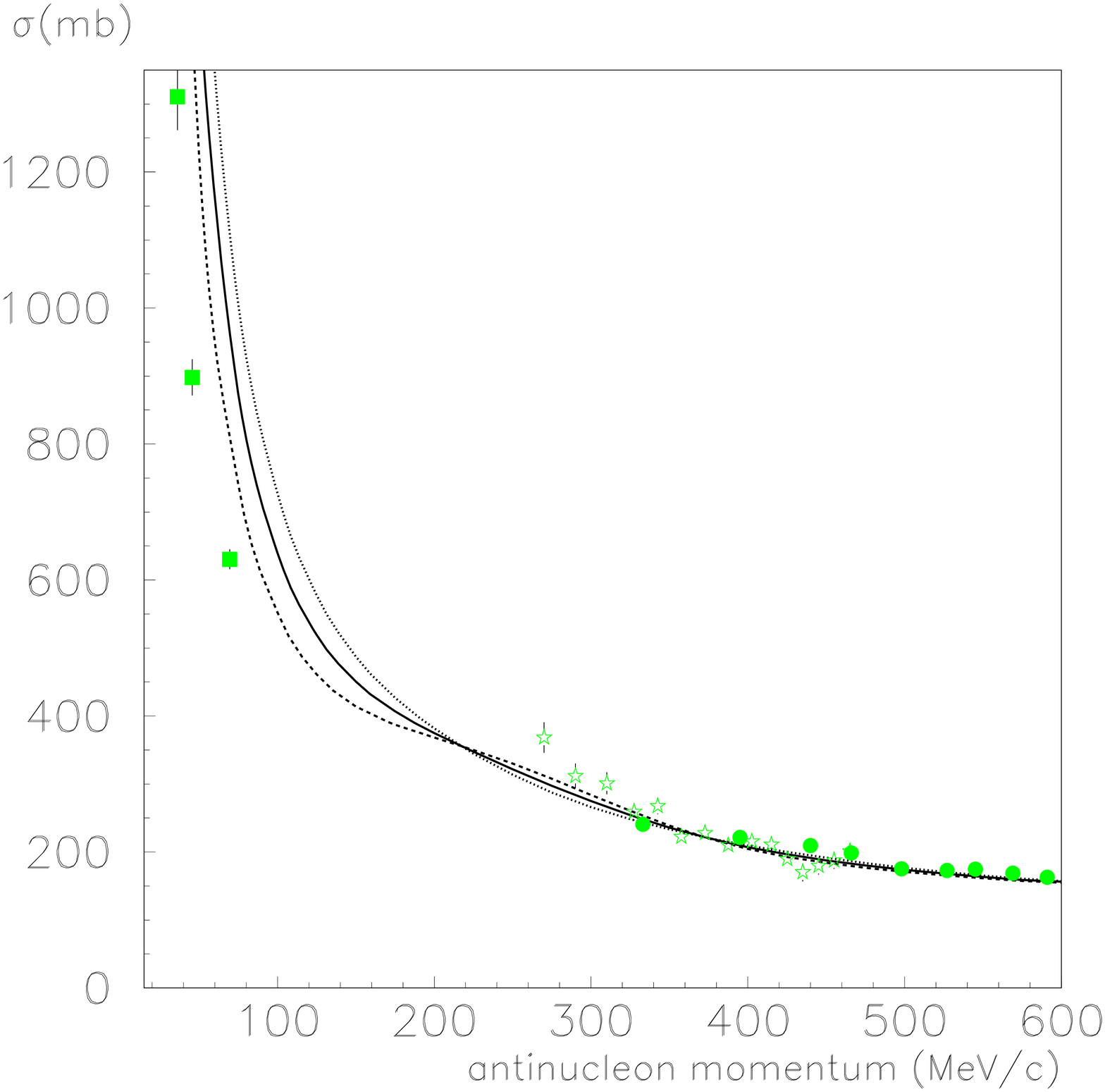,width=0.9\linewidth}}
\end{center}
\caption[]
{\small\it \label{fig5}
PWIA calculation of the $\bar{p}$-Deuteron total 
annihilation cross sections, together with data points 
taken from references \cite{obe2} (full squares), 
\cite{kalo80} (empty stars), \cite{bizz80} (full circles). 
Continuous 
line: full PWIA calculation with Coulomb correction and 
renormalization of the curve to the point at 340 MeV/c. 
Dashed line: the Deuteron is supposed to be composed 
by two neutrons (with overall nuclear charge Z=1). 
Dotted line: the Deuteron is supposed to be composed 
by two protons (with overall nuclear charge Z=1).
See the text for more details. 
}
\end{figure} 

\begin{figure}[htp]
\begin{center}
\mbox{
\epsfig{file=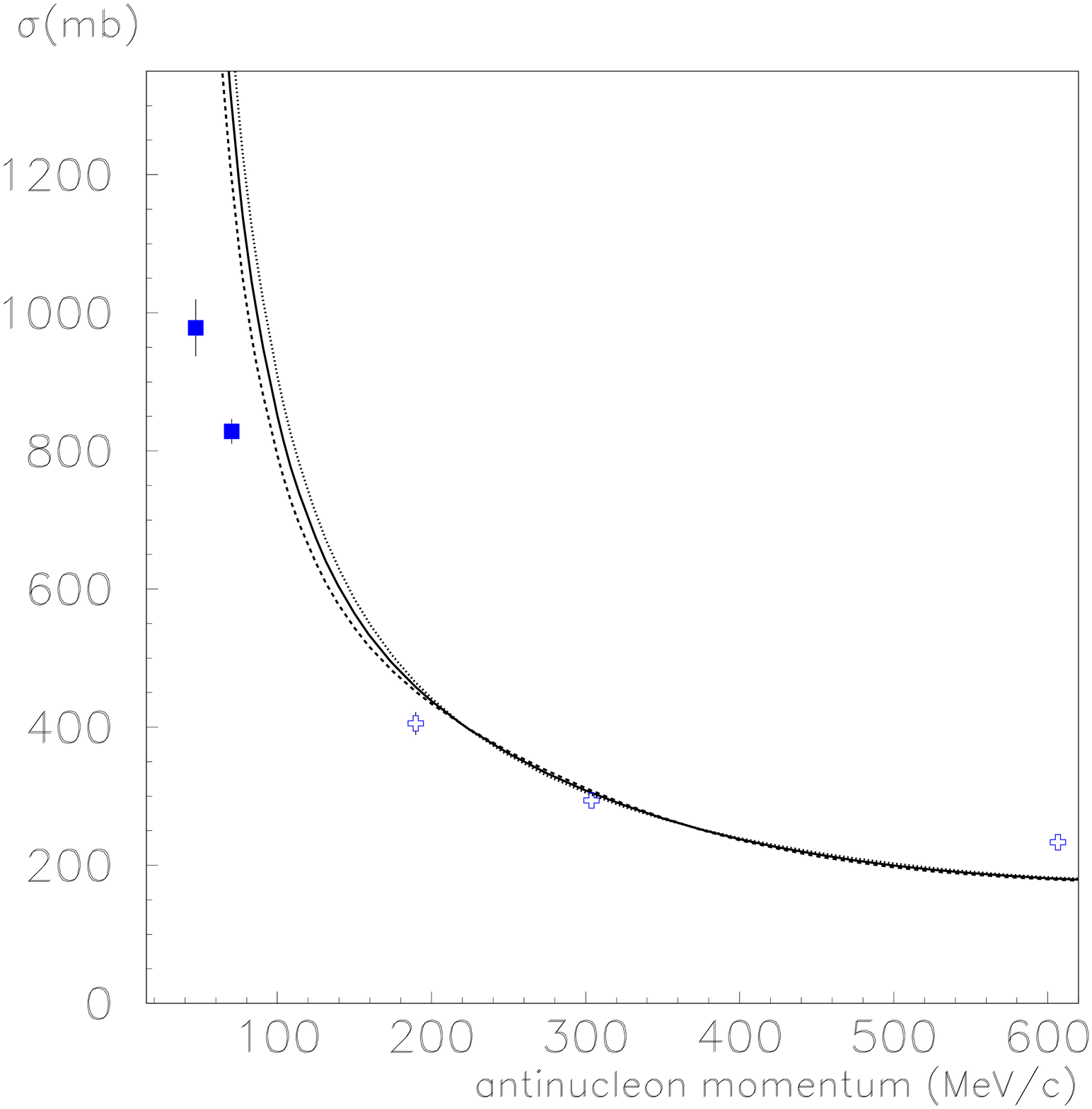,width=0.9\linewidth}}
\end{center}
\caption[]
{\small\it \label{fig6} 
PWIA calculation of the $\bar{p}$-$^4$He total 
annihilation cross sections, together with data points 
taken from references \cite{obe2} (full squares), 
\cite{bale85} (empty crosses). 
Continuous 
line: full PWIA calculation with Coulomb correction and 
renormalization of the curve to the point at 300 MeV/c. 
Dashed line: the nucleus is supposed to be composed 
by four neutrons (with overall nuclear charge Z=2). 
Dotted line: the Deuteron is supposed to be composed 
by four protons (with overall nuclear charge Z=2). 
See the text for more details. 
}
\end{figure}

In the last two figures we show some PWIA fits of the 
$\bar{p}$ annihilation cross sections on Deuteron and 
$^4$He. To obtain these curves the same above method 
has been used, but the final result has been multiplied, 
to take the nuclear charge into account 
in agreement with ref.\cite{n2}, by the factor 
$1 + Z C \beta^{-1.4}$, where $Z$ is the nuclear charge, 
$\beta$ is the $\bar{p}$-nucleus relative velocity 
and the constant $C$ is characteristic of the target nucleus. 
The exponent ($-$1.4) has no special physical meaning, since 
the above $1 + Z C \beta^{-1.4}$ factor is 
only a fitting relation that reproduces within some 
percent the Coulomb enhancement factor in the range 
40-400 MeV/c. The constant $C$ is 0.0060 for Deuteron, 
and 0.0040 for $^4He$. To take surface absorption 
into account, the two PWIA fits have been renormalized by 
a constant factor, the best to reproduce the data 
(or some data) over 200 MeV/c. We remark that: (i) 
this Coulomb factor is not relevant from 100 MeV/c onward, 
(ii) it is not very different between light nuclei 
at the same $\bar{p}$ momentum in the laboratory frame, (iii) 
in Hydrogen, Deuteron and $^4$He it is near 1.5 at 50 MeV/c 
and 2 at 25 MeV/c (laboratory frame). 

In the two figures we also report what would 
be the PWIA distribution if the 
target nucleus were composed by neutrons only or by protons 
only (without touching the Deuteron and $^4$He charges, 
which are taken into account by the $1 + Z C \beta^{-1.4}$ 
factor in any case). The first evident observation 
is that the effect of the $\bar{n}p$ gap/peak structure 
is much more evident in the Deuteron case. This is due to 
the compactness of the $^4$He structure, and to the 
exactly opposite peculiarity of the Deuteron structure.  
A more compact space structure implies a broader momentum 
distribution. The Fermi motion tends to average out the 
details of the momentum dependence of the nucleon-antinucleon 
annihilation rates. This mechanism is more effective with 
broader nuclear momentum distributions. 
Another observation is that, if we define shadowing as 
the departure between the observed data and 
Impulse Approximation predictions, 
this departure becomes evident in both cases below 100 MeV/c,  
but more evident in the $^4$He case. In the 
Deuteron case the shadowing phenomenon is less dramatic, 
but would be overestimated by not taking into account that 
one of the two nucleons composing this nucleus is a neutron. 

Concluding this section, we may say that some evident 
correlation exists between nuclear data and the behavior 
of the $\bar{n}p$ annihilation rate. In particular, it 
would be difficult to justify the shape of the $F_4$ and 
$F_2$ ratios in absence of the gap/peak structure of the 
free $\bar{n}p/\bar{p}p$ ratio. There is some chance that 
the shoulder found in the $\bar{n}$-nucleus annihilation 
cross sections is related with the same gap/peak structure, 
however in this case the situation is much less clear. 
Still to be clarified is the fact that the energy-averaged 
value of the $F_A$ ratio decreases at increasing $A$. 
The last two figures also show that the energy dependence of 
the nuclear annihilation rates can be partly explained in 
terms of the energy dependence of the $\bar{p}$ annihilation 
cross section on a free neutron. A consistent part of 
this energy dependence, however, cannot be explained in 
these terms and we may say that a strong 
energy-dependent shadowing is present. 

\section{Conclusions.}

We have presented an analysis comparing low energy 
$\bar{n}p$ annihilation data with other available 
data on antinucleon-nucleon and antinucleon-nucleus 
annihilation. The different sets are all consistent with 
a gap/peak structure in the isospin-1 channel, over a 
regular ``background'' which can be 
reproduced by energy-independent optical potentials of 
simple form. This structure is not too pronounced but 
evident, and we have also shown that it 
affects the nuclear annihilation rates, whose behavior 
can be $partly$ explained in terms of the behavior of the 
antiproton-neutron annihilation rate. 

We have also discussed 
the shadowing phenomena related with antinucleon-nucleus 
annihilation, and reported an interpretation in 
terms of the quantum uncertainty principle. 
The comparison of impulse approximation predictions with the 
nuclear data suggests that a strong energy-dependent 
shadowing is present at low energies: also after 
including absorption corrections, the data can not be 
fully explained in 
terms of individual antinucleon-nucleon annihilations.

\end{document}